# A hybrid privacy protection scheme for medical data

Judy X Yang, Hui Tian, Alan Wee-Chung Liew  and Ernest Foo

School of Information and Communication

Griffith University

Gold Coast, 4215, Australia

Email: Judy. yang@griffithuni.edu.au

Email: hui. tian@griffith.edu.au

Email: a. liew@griffith.edu.au

e.foo@griffith.edu.au

## ABSTRACT

Healthcare data contains patients' sensitive information, and it is challenging to persuade healthcare data owners to share their information for research purposes without any privacy assurance. The proposed hybrid medical data privacy protection scheme explores the possibility of providing adaptive privacy protection and data utility levels. The evaluation result demonstrates that the scheme can provide adaptive privacy and data utility levels, and the data holder can choose their preferred risk level and data utility through the scheme. Specifically, the data holders can assess their data attributes risk level based on mutual information score, which assures them to decide the privacy protection and data utility level. Data utility levels are evaluated based on the classification results after privacy preservation computing on healthcare data. Furthermore, Disclosure risk is evaluated based on the mutual information score of six types of healthcare data. The evaluation results on the heart disease and diabetes data demonstrate the scheme can provide a wide range of adaptive privacy protection and data utility levels to meet different privacy protection and data utility requirements.

## 1. Introduction

In the big data era, various cyberattacks increase the concerns of data owners about individual privacy disclosure. Especially in the healthcare domain, the rate of healthcare data leakage was 37%, which ranked the top among all the industries in past years. The second place was a retail business, at 10.9%. Healthcare data breaches spiked 55% in 2020. According to the CBS report, hackers can benefit at least 500 dollars per record. As healthcare data contains personal information, such as social security numbers, financial information, date of birth, medical record number, and insurance information, most hospitals and healthcare organisations are reluctant to share their data for research purposes [1].

Although the HIPAA (Health Insurance Portability and Accountability Act) [2] and GDPR (General Data Protection Regulation) Privacy Rule [3] have regulated privacy preservation related to individual information and Data Safe Harbour, health data leakage in the past years is still a significant concern for many data owners. Therefore, seeking suitable privacy-preserving technology is essential to reduce the concerns of data holders before data publishing.

Several anonymisation technologies apply to privacy protection, which is Naïve ID Removal, Anonymity by randomization, Anonymity by indistinguishability [4],  Differential Privacy(DP) and Local differential privacy (LDP) based techniques [5], Statistical Disclosure Control (SDC) [6], and Data encryption [7]. These techniques are used to preserve medical data privacy by perturbing the records with data utility in consideration.

Based on the basic k-anonymity model proposed by Sweeney [8], many extended k-anonymity, l-diversity, and t-closeness have been added to the model to defend against different attacks, e.g., background knowledge attacks [9], homogenous attacks [10], linkage attacks [11], unsorted matching attacks [12], complementary attacks and time-series inference attacks [13]. By developing different models, researchers seek a trade-off between privacy protection and data utility levels in scenarios with various attacks.

Federated learning (FL) is a distributed machine learning (ML) technique that can ensure collaborative training is performed using local datasets to protect the data owners' privacy. FL works as a collaborative process where edge devices train the model locally, and only the weight and gradient parameters need to be shared with the central server. However, the main drawback of the FL is that it cannot have additional data to verify its model effectiveness locally [14]. Therefore, it is challenging for FL users to access validation data from the FL architecture.

When designing a privacy protection model, the trade-off between privacy and data utility levels must be considered. It also requires quantifying the risks of privacy violations when designing a scheme. SDC and PPDP (Privacy-preserving data publishing) are two widely used privacy protection technologies. PPDP can be categorised into Randomisation and indistinguishability, and SDC is divided into masking techniques and synthetic data generation.

Researchers expect the published data can retain their required attributes to facilitate analysis. In [15], Hyukki Lee and his colleagues built a utility and anonymisation model for health data publishing. Their main idea is to insert counterfeit records in the data to preserve privacy. In addition, the utility error rate is used as a performance metric.

Authors in [16] proposed a fixed interval approach for medical data privacy with categorical data. In his paper, Natalie Shlomo introduces microdata disclosure risk estimation, data masking, and utility assessment.

Although many complex anonymisation techniques and de-identifying models have been developed, some excessive privacy preservation methods result in published data with a low utility level. Therefore, we propose a hybrid data protection scheme to improve such problems by providing adaptive privacy protection and utility levels. We compare different techniques, including k-anonymity, (k,l)-anonymity, randomization, Secure Lookup Masking, Binning and their hybrid methods, and study their privacy protection and data utility levels. Finally, an adaptive hybrid model is proposed to meet different privacy protection and data utility requirements.

Our work presented in this paper differs from the related work in several aspects:

First, instead of preserving as much data utility as possible, our research proposes a hybrid scheme that can provide adaptive privacy protection and data utility. The data holder can choose their preferred risk level and data utility level through a hybrid privacy protection method.

Second, we study the impact of our hybrid data privacy-preserving method on medical data by quantifying how much data information is preserved after privacy protection and how much disclosure risk is reduced.

Finally, we study the impact of the hybrid data privacy-preserving method on the accuracy of published data classification.

The main contribution of this paper can be summarised as follow: (1) A hybrid privacy preservation scheme is proposed for healthcare data. The result demonstrates that the proposed hybrid methods can achieve an adaptive range of data utilities and disclosure risk levels to meet publishing requirements. (2) The scheme provides a choice for data holders and lets them know their data disclosure risk level before and after data is published. Furthermore, by studying the ideas of existing technologies, the algorithms from two categories of privacy protection data publish (PPDP) techniques and Statistical Disclosure Control (SDC) are chosen to defend against the majority of available attacks and retain data utility based on the requirements of both data holder and data users. (3) The experimental results using a heart disease dataset and diabetes dataset demonstrate that the scheme is robust. Therefore, we believe the proposed method can be an essential reference tool for data publishing and analysis.

The rest of the paper is organised as follows: Related work is introduced in Section 2. Then, section 3 introduces the relevant concepts, and section 4 introduces five privacy-preserving methods applicable to healthcare data and evaluation. Finally, section 5 concludes the paper and outlines the future research plan.

## 2. Analysis of Hybrid Medical Data Privacy Protection

This section describes the HMDPP scheme, including medical data attributes, architecture, applied algorithms, utility assessment and disclosure risk level assessment.

### 2.1. Medical Data Attributes

Medical data management follows the General Data Protection Regulation (GDPR), the strictest privacy and security law globally. According to GDPR, Structured medical data includes patient personal information, like Diagnoses, Drugs and Procedures. This medical information can be divided into four types based on the attributes below.

Direct Identity Attributes (ID) include name, social security number, medical number, driver's license, Telephone number, etc.

Quasi-identifier Attributes (QID) include address, birth date, age, height, weight, gender, eye colour, hair colour, job, education, etc.

Sensitive Attributes (S) include health status, racial or ethnic origin, political opinions, religious or philosophical beliefs, trade union membership, genetic data, income, hobbies, family disease history, marital status, sexual orientation, etc.

Non-Sensitive attributes. (NoS) includes Admission time, discharge time, Radiology data, medical lab test data, medical image data and others.

Conventionally, the ID attributes of the medical data have been deleted. This paper defines the original dataset as $A$ without ID attributes already. The subset dataset $Q$ represents Quasi-identifier attribute data. $S$ is the sub-data set of sensitive attributes; $M$ is the non-sensitive dataset with non-sensitive attributes; $Y$ is the diagnosis result. Therefore, the original dataset can be expressed as $A = Q \cup S \cup M \cup Y$

The original dataset is partitioned into the sub-datasets first, and then privacy protection is applied to the subsets $Q$ and $S$, and then $Q'$ and $S'$ are obtained, while $M$ and $Y$ are retained without any privacy protection. Therefore, the published dataset can be expressed as $A' = Q' \cup S' \cup M \cup Y$.

### 2.2. HMDPP System Description

Figure 1 illustrates the conceptual workflow of the Hybrid Medical Data Privacy Protection (HMDPP) method for adaptive privacy protection and data utility level.

The mutual information between attributes and diagnosis are calculated and compared based on $A$ and $A'$. The classification performance is compared between $A$ $and$ $A'$ as well.

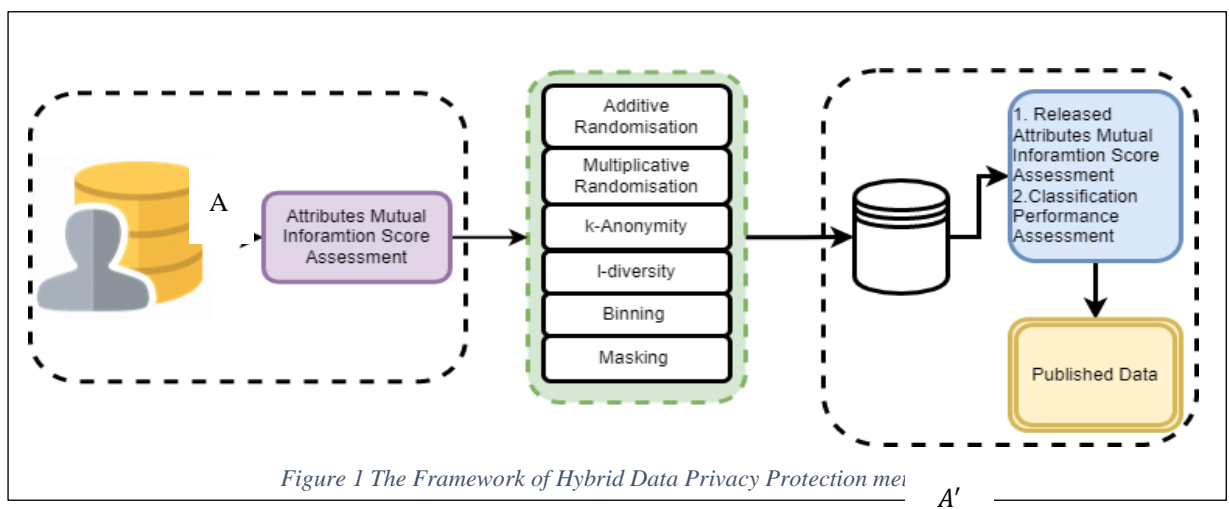

*Figure 1 The Framework of Hybrid Data Privacy Protection me...*

The system data flow can be described as follows,

**A**. Original data partition and attributes assessment: The data owner uses this module to assess attribute disclosure risk levels. First, it partitions the medical data table into three sections: *Q, S, M* and *Y*. After this step, mutual information is calculated between the Target (*Y*) and attributes, and then the attributes disclosure risk level is provided to the data holder for assessment.

**B**. Hybrid Privacy Protection Methods Applied: The data user utilises this module to implement their medical data privacy protection based on their preferences.

**C**. Released data assessment: the data owner can utilise this module to assess their data disclosure risk level and data utility through classification.

By selecting an appropriate method for protecting data privacy before publishing, the system can provide adaptive privacy protection and data utility level based on the requirements of both data holders and data users.

### 2.3. Algorithms Applied in HMDPP

According to the research in [17], data privacy protection techniques can be applied to medical microdata and medical databases. Considering that our proposed scheme is intended for tabular medical data, the chosen algorithms in Figure 2 are from PPDP and SDC.

The Privacy Protection Data Publishing (PPDP) or Privacy Protection Data Management (PPDM) category can defend against homogeneity attacks, background knowledge attacks, and query-based attacks, whereas Statistical Disclosure Control (SDC) can defend against inferred linking attacks. Thus, six privacy protection methods, consisting of techniques from two categories, are chosen to implement privacy protection for quasi-identifiers (QID) and sensitive attributes. Such selection can assure that these privacy protection methods can defend against most attacks.

Both Neural Networks and Database privacy technologies will be our future research direction.

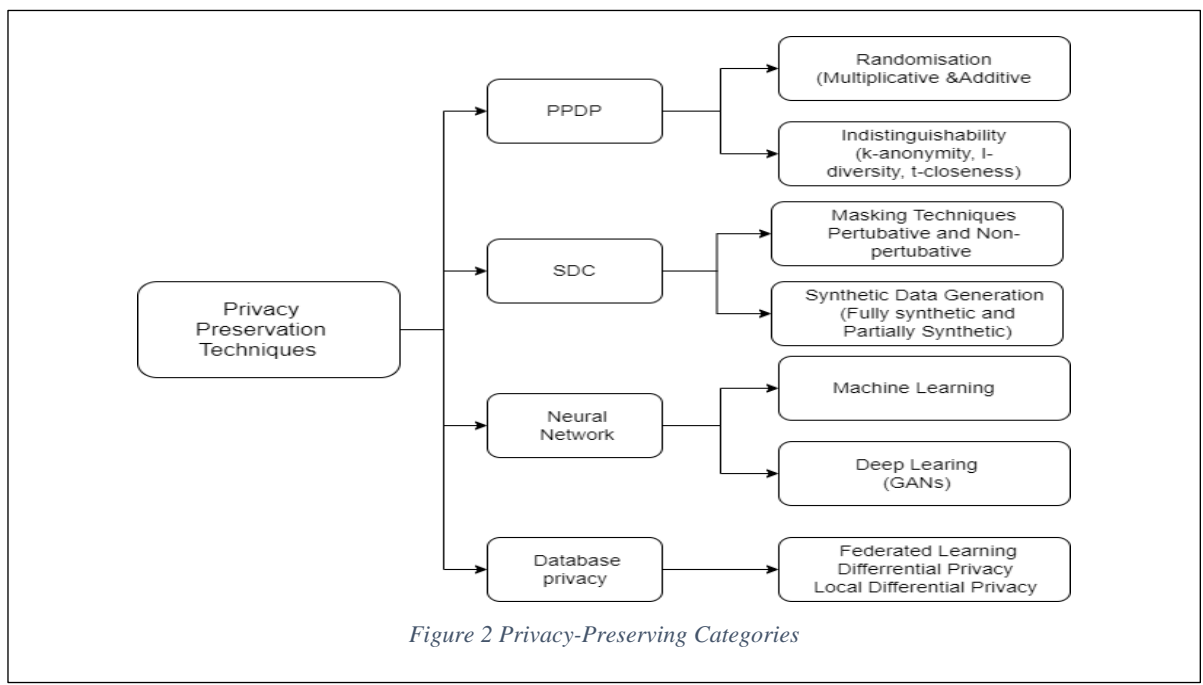

*Figure 2 Privacy-Preserving Categories*

The six algorithms are briefly described as follows,

**Randomisation Additive (RA)** [17]**:** The original dataset is denoted by $X= Q \cup S \cup M$, and $X=\{X^1, X^2, ..., X^m\}$ represents the original elements in the dataset. The new privacy protection dataset, denoted by $Z= Q' \cup S' \cup M'$, and $Z=\{Z^1, Z^2, ..., Z^m\}$ is acquired through adding normal distribution noise to each record.

**Randomisation Multiplicative (RM)** [17]**:** Similarly to Randomisation additive, the original dataset is denoted by $X= Q \cup S \cup M$, and $X=\{X^1, X^2, ..., X^m\}$ represents the original dataset. The new privacy protection dataset, denoted by Z= $Q' \cup S' \cup M'$, and Z= $\{Z^1, Z^2, ..., Z^m\}$ is acquired through adding Gaussian distribution noise to each record.

**K-anonymity (k.)** [17]: The property of k-anonymity is that the released record has at least (k-1) other records shown in the released data, which can defend against record linkage. Generalization and suppression are usually employed to reduce the granularity of representation of quasi-identifiers or QID ($Q$)

**l-diversity (l.)** [18]: *l*-diversity was proposed to defend against the limitation of k-anonymity. As an extension to k-anonymity, it is used to overcome homogeneity and background knowledge attacks.

**Binning** [19]: Data binning, also known as discretisation, categorisation or quantisation, simplifies and compresses a column of data by reducing the number of possible values or levels represented in the data.

**Masking** [20]: The original data are transformed to produce new data that are valid for statistical analysis and such that they preserve the confidentiality of respondents.

Replacing sensitive values with other characters like "*" is also known as data masking.

The chosen six algorithms can be selected separately or combined to perform privacy protection based on the user's preference.

In addition, five hybrid methods are given by a combination of the six algorithms, i.e., *(RA, k BIN, MASK), (BIN, MASK), (k, l), (RA), and (RM)*, which are also implemented in the system.

Based on the mutual information score of the medical data attributes, the data holders can choose their preferred privacy protection methods to meet their requirements of privacy protection and data utility. As illustrated in Figure 2, the proposed solution involves three different steps: (a) Data holders check their data $Q$ and sensitive attributes mutual information score, which can assist them in selecting their privacy protection method; (b) Implementing hybrid privacy protection methods based on the importance of their attribute; (c) Assessing the classification performance and the $Q'$ and sensitive attributes information score after privacy protection.

The abbreviations of the chosen six privacy protection algorithms are shown in Table 1.

*Table 1 Algorithm Abbreviations*

| Item | Selected Algorithm | Algorithms Abbreviation |
|---|---|---|
| 1 | Randomisation Additive | RA |
| 2 | Randomisation Multiplicative | RM |
| 3 | k-anonymity | k |
| 4 | l-diversity | l |
| 5 | Binning | BIN |
| 6 | Masking | MASK |

### 2.4. Utility Analysis

The *Accuracy, Precision, Recall* and *f1-score* measure the classification performance of the original and published data with different privacy protection methods. They are defined as follows.

$$Accuracy = \frac{TP+TN}{TP+TN+FP+FN} \quad (1)$$

$$Precision = \frac{TP}{TP+FP} \quad (2)$$

$$Recall = \frac{TP}{TP+FN} \quad (3)$$

$$f1 - score = 2 * \frac{precision*recall}{precision+recall} \quad (4)$$

In the experiments, the classification results of the raw data and the other five data with privacy protection will be compared based on equations (1) to (4).

### 2.5. Disclosure Risk Analysis

The techniques are proposed for medical microdata. However, such data are not always suitable for data with a complex nature. Therefore, in this paper, the direct information of patients has been removed, and the quasi-identifier data, sensitive medical data, common clinical data, and Target are retained in the table for further privacy protection.

Let A denotes the set of all attributes and targets $\{X^1, X^2, ..., X^m: Y\}$.

We quantify the disclosure risk using information theory. Specifically, we measure the disclosure risk by calculating the mutual information (*MI*) [21] between attributes $X^i$ and $Y$. For attribute $x \in X^i$ and $y \in Y$, *MI* is defined as

$$MI(X^i:Y) = \int dxdy(p(x,y) \log \frac{p(x,y)}{P_X(x)P_Y(y)}) \quad (5)$$

The *MI* determines the attribute's disclosure risk level. The higher the value, the more information is contained in the attribute, which is easy for the adversaries to perform re-identification.

Besides, the dataset's overall *MI* can be used to assess the data disclosure risk reduced ratio between the raw data and the published data. It can be expressed by the equation as follow,

$$Risk_{reduced} = \frac{(\sum_{X \in A} MI(X;Y) - \sum_{X' \in A'} MI(X';Y))}{\sum_{X \in A} MI(X;Y)} \quad (6)$$

Where $A$ is the raw data, $A'$ is the published data.

## 3. System Evaluation

The HMDPP results are presented in this section based on the *Heart Disease Data* and *Diabetes Diagnosis Data.*

The two disease records sourced from Kaggle have been pre-processed. First, the raw data is divided into four partitions $\{Q, S, M, Y\}$, $Q$ is a quasi-identifier subset, $S$ is the subset of the sensitive attributes, $M$ is the common records, and $Y$ is the diagnosis result. Specifically, five privacy protection methods, i.e., (*BIN, MASK*), (*k, l*), *RM, RA, (RA, k, BIN, MASK),* are implemented in the experiment to evaluate the scheme's performance. Then, the five data performance with privacy protection for data utility evaluation is compared with the original dataset. In addition, the classification performance is evaluated through KNN Logistic Regression, and disclosure risk is evaluated using mutual information based on raw and published data.

All experiments are conducted on a standard PC, x64-based system, with 8GB of RAM. Google Colab is used to run the program.

### 3.1. Utility Evaluation

This subsection evaluates the utility performance of the privacy protection algorithms based on two data of "heart disease" and "Diabetes".

We have four cases in the heart disease and diabetes data:

True positive (TP): the observed case is 1 (having heart disease or diabetes) in the data that the classifier prediction is 1 (having heart disease or having diabetes).

True negatives (TN): the observed case is 0 (No heart disease or No diabetes) in the data that the classifier prediction is 0 ('No heart disease or No diabetes).

False positives (FP): the observed case is 0 (No heart disease or No diabetes) in the data that the classifier prediction is 1 (Having heart disease or Having diabetes).

False negatives (FN): the observed case is 1 (Having heart disease or Having diabetes), and the classifier prediction is 0 (No heart disease or No diabetes).

Take the output of the raw data classifier confusion matrix, and we can interpret the outcome performance.

For heart disease data, RA and MA are used to add noise to the whole data through normal distribution; (BIN, MASK) is applied to attributes *Age* and *Thalach*; (k,l) is applied to *Age* and *CP*; (*RA, k, BIN, MASK*) is applied to the whole data by *RA,* and then (*k-anonymity)* for *Age,* (*BIN, MASK*) is applied to *CP* and *CA.*

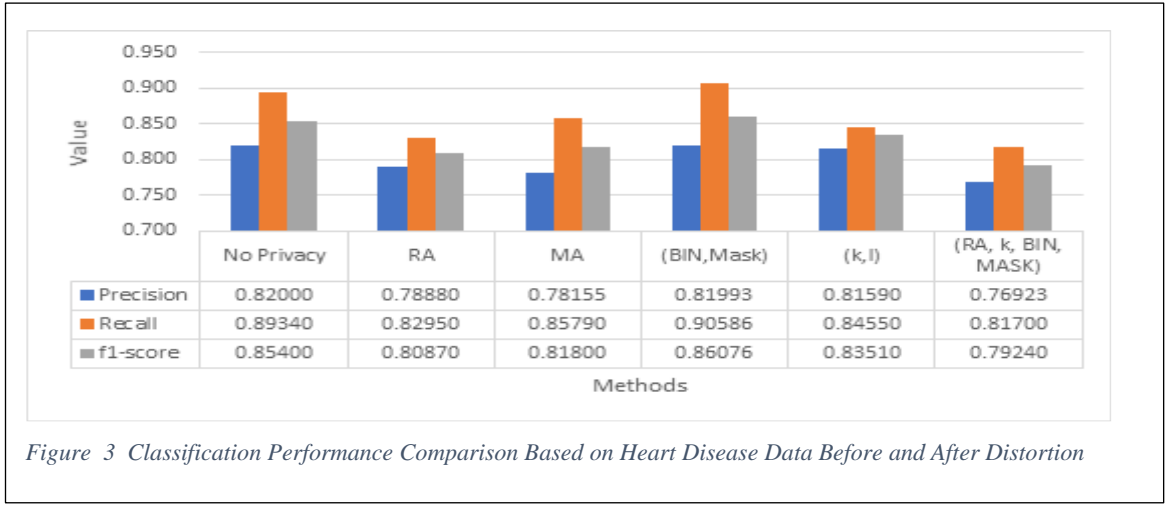

*Figure 3 Classification Performance Comparison Based on Heart Disease Data Before and After Distortion*

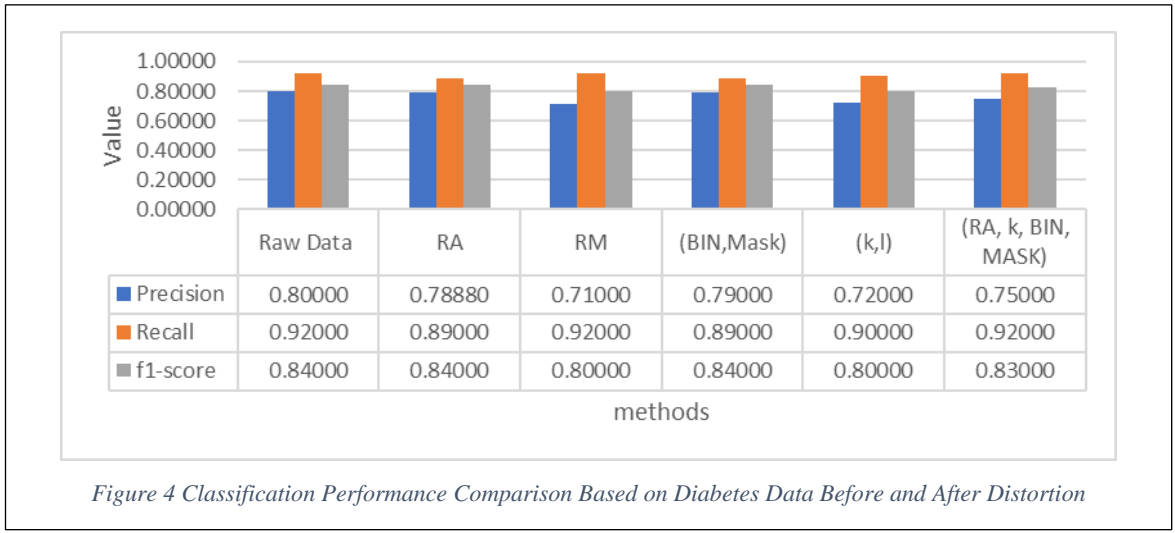

*Figure 4 Classification Performance Comparison Based on Diabetes Data Before and After Distortion*

Figure 3 shows the heart disease classification performance based on data before and after distortion. The outcome shows that the different results of classification precision adaptive range from 0.6%~5% and Recall adaptive range is from 0%-7%, the f1-score adaptive range is from 4%-6%. Notably, we demonstrate that the data with the hybrid method (*BIN, MASK*) has a similar classification performance to the raw data without privacy protection. In contrast, the other hybrid method test result shows the difference is a bit large but still within 5% in terms of Precision. Therefore, the data holders can choose their preference classification performance based on their preference or the negotiation results with the data users.

While for diabetes data utility evaluation, the same privacy protection methods are applied to the data also. *RA* and *MA* are used to add noise to the whole data through normal distribution; (*BIN, MASK*) is applied to attributes *Glucose* and *Age*; (*k, l*) is applied to *Age* and *BMI*; (*RA, k, BIN, MASK*) is applied to the whole data by *RA* first, and then (k) for *Age*, (*BIN, MASK*) is applied to *BMI* and *Glucose*.

Figure 4 shows the diabetes classification performance based on data before and after distortion. The result shows that the Precision adaptive range is from 0 to 5%, and both Recall and f1-score are nearly at the same level compared with the raw data.

The two experimental results based on two data types show that the data utility can remain at a higher level even after adding privacy protections to the raw data.

### 3.2. Disclosure Risk Evaluation

This subsection evaluates the system defends against attacks and the relationship between classification performance and attributes mutual information change.

We calculate the *MI* score of attributes on the six data sets to investigate some critical attributes information scores change before and after the privacy protection is applied. Equation (5) is applied to calculate attributes *MI*.

Figure 5 shows the heart disease attributes of Mutual Information for six types of datasets before and after the privacy protection methods were applied. Equation 5 illustrates the Mutual information calculation process. The hybrid algorithm (*BIN, MASK*) can cause the data leakage information at the minimum level. In contrast, the hybrid method (RA, k, BIN, MASK) can cause information loss at the maximum level. The maximum information loss rate is 72% compared with raw data.

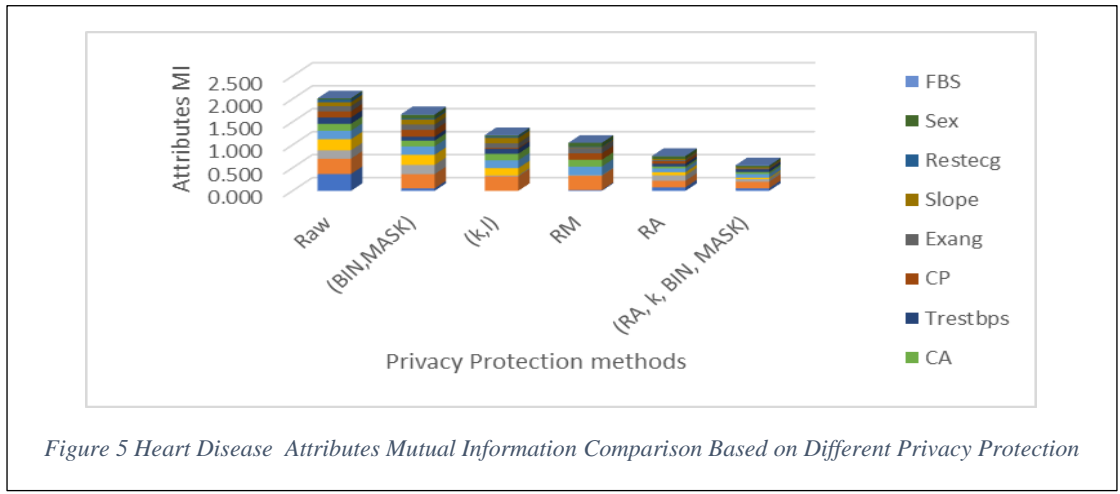

*Figure 5 Heart Disease Attributes Mutual Information Comparison Based on Different Privacy Protection*

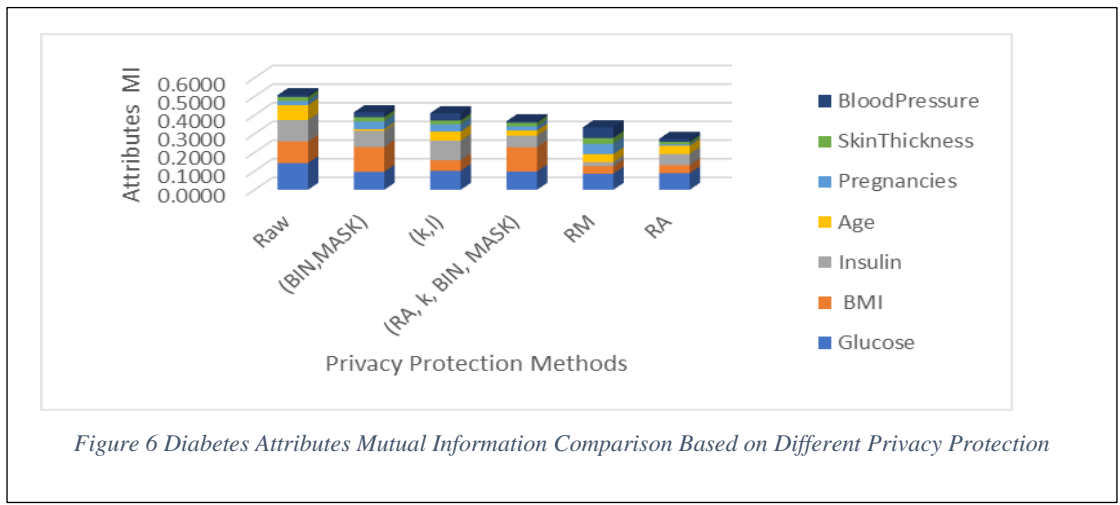

*Figure 6 Diabetes Attributes Mutual Information Comparison Based on Different Privacy Protection*

Figure 6 shows the diabetes disease attributes of Mutual Information for six types of datasets before and after the privacy protection methods were applied. Like heart disease data test results, (*BIN, MASK*) can cause data information leakage at the minimum level. In contrast, the hybrid method (**RA, k, BIN, MASK**) can cause data information leakage at the maximum level. The MI loss rate is 46% compared with the raw data.

### 3.3. Results Analysis

We evaluate the proposed HMDPP scheme in sections 4.1, 4.2 and 4.3. Due to the space limitation, we have five hybrid algorithms to evaluate the utility performance, disclosure risk and de-attack based on heart disease and diabetes data. The experimental results show that the heart disease classification adaptive ranges are 1% to 6% based on the raw data and data with privacy protections, whereas the diabetes classification precision adaptive range is 1% to 5%. However, the data holder can choose their preference privacy protection and utility level through the HMDPP system.

The reasonable privacy protection method can be achieved by reducing the attributes information regardless of adversaries' background knowledge. Section 4.2 evaluates disclosure risk reduced ratios based on mutual information. According to the evaluation result, the different privacy protection methods can result in different attributes mutual information reduced ratios, and the corresponding attributes disclosure risk will be reduced accordingly. For example, the heart disease total mutual information reduced rate changes range is from 17% to 72%, and diabetes total mutual information change range is from 18%-46%. The process can quantify disclosure risk for data holders and let them know their data disclosure risk level.

Finally, the hybrid (BIN, MASK) has the best classification precision but less privacy protection, whereas the Hybrid (BIN, MASK, k, MA) has maximum information loss and utility loss.

## 4. Conclusions and future work

This section summarises the system evaluation results, contributions, and future research plan.

No privacy protection scheme can be optimal to achieve both the best privacy protection and maximum data utility. Therefore, this paper proposes the HMDPP scheme for medical data privacy protection, which can provide adaptive privacy protection and data utility level based on data holders' preferences.

Five data privacy-preserving methods are selected to implement system evaluation. The experiment results demonstrate the robustness of the HMDPP scheme on both adaptive privacy protection and data utility. First, the data holder can be advised on their data privacy level based on mutual information score before they have the intention to publish their data through the HMDPP system. The HMDPP method can reduce data risk levels by 72.25% on heart disease data and 46% on diabetes data, and the classification precision reduced range is 6%. Besides, the classification performance shows that the system can provide adaptive privacy protection and data utility level based on the preference of the data holder. Finally, the disclosure risk level is assessed by using mutual information score based on datasets before and after distortion.

Our contributions to this paper can be summarised as follows:

First, we propose a hybrid system to ensure that data privacy protection can provide adaptive privacy protection and utility level. A series of experiments have demonstrated that our methods are effective and can provide practically acceptable values for balancing privacy and accuracy in medical data classification.

Secondly, the HMDPP scheme ensures the data holder knows about the data attributes risk level before and after privacy protection. Furthermore, the proposed HMDPP methods allow the data holder to choose their preference privacy protection and data utility level.

Finally, the HMDPP scheme can assess the classification performance of the published data, which is valuable information for both data holders and data users.

The results of the experiments demonstrate that our methods can achieve expected results and are promising. Currently, we are extending our work in the following directions: (a) we are planning to fine-tune the interface system based on attributes risk level selection for data holders; (b) Other medical data privacy methods, such as synthetic techniques combined with Generative Adversary Networks(GANs), will be added into the system to enhance the system ability; (c) Medical database privacy based on Federated Learning will be explored, differential privacy and local differential privacy, and a reward system combined with Federated learning will be explored also.